\documentclass[twocolumn,tighten]{aastex63}
\pdfoutput=1 
\usepackage{amsmath,amstext}
\usepackage{apjfonts}
\usepackage{afterpage}
\usepackage{xcolor}
\usepackage{threeparttable}
\usepackage{lineno}
\usepackage{graphicx}
\usepackage{epsfig}
\usepackage{booktabs}
\usepackage{multirow}
\usepackage{float}
\usepackage{mathrsfs}
\DeclareUnicodeCharacter{1E11}{\k{a}}
\setwatermarkfontsize{100px}

\begin{document}
\author{Tong Zhao}

\affiliation{School of Physics, Georgia Institute of Technology, 837 State St NW, Atlanta, GA 30332, USA}
\affiliation{Departments of Astronomy and Physics, University of Arizona, 933 N. Cherry Ave., Tucson, AZ 85721, USA}

\author[0000-0003-1035-3240]{Dimitrios Psaltis}
\affiliation{School of Physics, Georgia Institute of Technology, 837 State St NW, Atlanta, GA 30332, USA}

\author[0000-0003-4413-1523]{Feryal \"{O}zel}
\affiliation{School of Physics, Georgia Institute of Technology, 837 State St NW, Atlanta, GA 30332, USA}

\author{Elif Beklen}
\affiliation{Department of Physics, Suleyman Demirel University, 32260, Isparta, Turkey}


\title{Uncovering Correlations and Biases in Parameter Inference from Neutron-Star Pulse Profile Modeling}

\begin{abstract}
   Modeling of X-ray pulse profiles from millisecond pulsars offers a promising method of inferring the mass-to-radius ratios of neutron stars. Recent observations with NICER resulted in measurements of radii for three neutron stars using this technique. In this paper, we explore correlations between model parameters and the degree to which individual parameters can be inferred from pulse profiles, using an analytic model that allows for an efficient and interpretable exploration. We introduce a new set of model parameters that reduce the most prominent correlations and allow for an efficient sampling of posteriors. We then demonstrate that the degree of beaming of radiation emerging from the neutron star surface has a large impact on the uncertainties in the inferred model parameters. Finally, we show that the uncertainties in the model parameters for neutron stars for which the polar cap temperature falls outside of the NICER energy range are significantly degraded. 

\end{abstract}

\section{Introduction}
Measurements of the neutron star mass-radius relation provide an indispensable way to constrain the equation of state of cold dense matter. While the mass of a neutron star in a binary system can be measured with high precision via radio timing \citep[see, e.g.,][]{Shapiro1,Shapiro2},  precise measurements of neutron star radii remain challenging~\citep[see][for a review]{Ozel2016b}. 

There are several approaches to measuring neutron star radii, either directly or indirectly. In earlier work, radii have been inferred from accreting neutron stars in quiescence~\citep{Heinke2003,Heinke2006} and from the spectra of Eddington-limited X-ray bursts~\citep{Ozel2009,Guver2010a,Guver2010b}, and the results have been used to place constraints on the equation of state of neutron-star matter~\citep{Ozel2010,Ozel2016a,Steiner2010,Steiner2013}. More recently, the detection of gravitational waves from neutron star merger events have placed constraints on the tidal deformability and, implicitly, on the radii of the two neutron stars in the coalescing system~\citep{GW,Raithel2018}.

A third avenue to measuring neutron-star masses and radii is through the analysis of X-ray pulse profiles. These include modeling the X-ray light curves of accretion-powered millisecond X-ray pulsars~\citep{accretion1,accretion2}, of millisecond neutron stars with short-lived oscillations during thermonuclear X-ray bursts~\citep{burst1,burst2,burst3}, and of rotation-powered millisecond pulsars that produce periodic X-ray emission from hot polar caps~\citep{thermX1,thermX2,thermX3}. 

For the last category, the Neutron Star Interior Composition Explorer (NICER) was designed primarily for observing several nearby millisecond X-ray pulsars and measuring radii and masses via fitting model pulse profiles~\citep{NICER1,NICER2,NICER3}. The initial application of this technique to observations of the millisecond pulsars PSR~J0030+0451 \citep{Miller2019,Riley2019} and PSR~J0740+6620~\citep{Miller2021,Riley2021} resulted in inferred radii that appear, at face value, to be larger than and, hence, in tension with those measured by the spectroscopic methods and inferred by the gravitational wave observations~\citep[see, e.g.,][]{Raithel2021}. Furthermore, modeling of their pulse profiles yielded complex geometries for the hot spots, with temperatures, sizes, and clustering that do not naturally arise from pulsar models. 

A follow-up analysis by \citet{Vinciguerra2023} revealed a number of vulnerabilities of the analysis methods due to biases introduced by the presence of significant measurement uncertainties and the inability of the numerical algorithm to sample the multimodal structure of the posteriors over the many model parameters. Indeed, reanalyses of the NICER observations of  PSR~J0030+0451 and PSR~J0740+6620~\citep{Vinciguerra2024,Salmi2024} resulted in revised measurements of the neutron-star radii in these sources to smaller values ($\sim 11$~km), aligning them with the earlier measurements based on Eddington-limited X-ray bursts~\citep{Ozel2016a}.  A more recent analysis of the pulse profiles observed with NICER from PSR~J0437$-$4715 also resulted in similarly small values for the inferred neutron-star radii~\citep{Choudhury2024}. The significant revision of the inferred neutron-star radii from pulse profile modeling raises questions about degeneracies between parameters and the completness of the parameter space exploration, thus warranting a careful investigation of degeneracies and biases in this method, which we explore in this paper.

The fundamental difficulty in these analyses arises from the fact that modeling the pulse profiles from spinning neutron stars involves relativistic ray-tracing from hot spots on the neutron star surface to an observer at infinity, which can be computed only numerically and is computationally expensive~\citep{Pechenick1983}. Such numerical approaches make it difficult to uncover degeneracies of the results on model parameters or identify the origin of potential biases (e.g., \citealt{miller2015,salmi2023}). This is especially true for neutron stars spinning at frequencies $\gtrsim 300$~Hz, the shapes and spacetimes of which are not spherically symmetric~\citep{Cadeau2007,Psaltis2014}.  In the regime of slow rotation, however, which is relevant to the NICER sources, there is an approximate analytic solution in the Schwarzschild+Doppler (S+D) limit~\citep{Pout1}, in which the external spacetime is described by the Schwarzschild metric, the spin of the neutron star introduces only Doppler effects, and the degree of gravitational light bending is modeled via an approximate analytic function. 

The simple analytic formula allows us to calculate the observed spectrum from a small spot on a neutron star, given a set of relevant parameters, such as the mass and radius of the neutron star, the spectrum of emission from its surface, as well as the orientation of the observer and the location of the hot spot. This approximation was used by, e.g.,~\citet{Ozel2016c} to investigate requirements on the prior knowledge of the NICER instrument backgrounds to achieve a target precision in the measurement of neutron-star radii. In this paper, we use this approximation to build a complete analytic model for pulse profile generation, incorporating the effects of the energy dependent beaming of radiation emerging from the neutron-star surface. We use this first to identify previously unexplored correlations between model parameters and define appropriate combinations of model parameters that reduce these correlations. We then explore the impact and potential bias of the limited photon energy range of the observations on the inference of neutron-star radii.

In \S2, we develop the analytic model and, in \S3, we devise a set of model parameters that reduce correlations and degeneracies between them. In \S4, we describe our approach to making synthetic data as well as approach to fitting these data via Markov-Chain Monte Carlo methods. In \S5, we explore a number of effects that limit the accuracy of the inference of model parameters and, in \S6 we provide a brief summary of our results.

\section{An Approximate Analytic Model for Pulse Profile Modeling}

In this section, we introduce our approximate analytic model of the neutron star thermal X-ray pulse profiles. The thermal X-ray emission is thought to arise from localized ``hotspots'' on the neutron-star surface that come in and out of view as the star rotates. The observed flux depends on a number of geometric parameters, such as the location and size of the spot(s), and the inclination of the observer. It also depends on the strength of gravitational lensing, as measured by the compactness of the neutron star, as well as on the temperature and the angular dependence of emission from the neutron-star atmosphere. Throughout the paper, we consider a neutron star of mass $M$ and radius $R$, such that its compactness is $u\equiv 2 GM/c^2$, where $G$ and $c$ are the gravitational constant and the speed of light, respectively.

\subsection{Geometry of the problem}

\begin{figure}[t]
	\centering
	\includegraphics[width=0.9\linewidth, keepaspectratio]{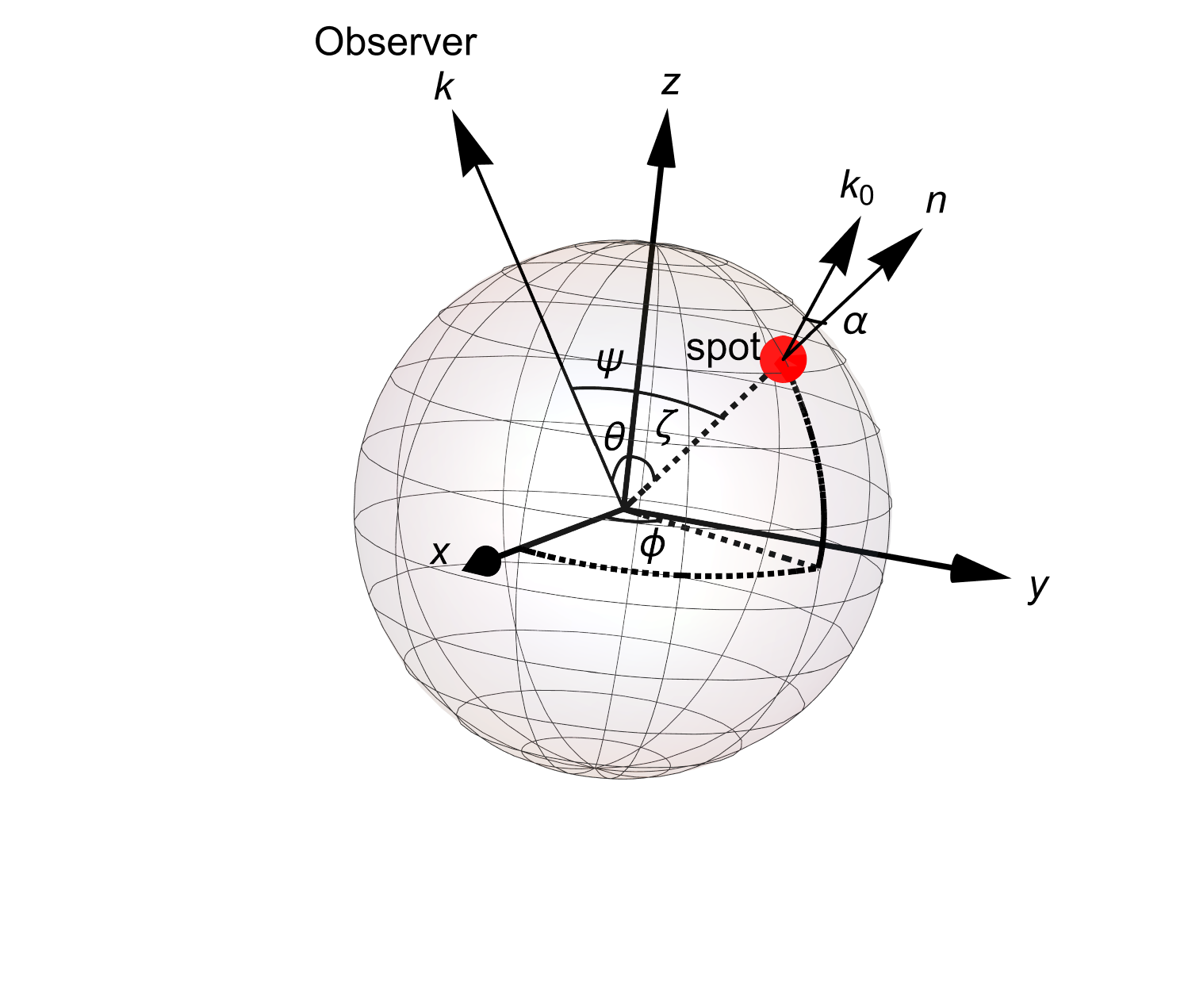}
	\caption{\label{fig:geometry} The geometry of localized emission on the surface of a neutron star that we consider in this paper. A hot spot is located on the surface at a colatitude $\zeta$ and a phase angle $\phi$. The observer is looking at the star center at an inclination angle $\theta$. The unit surface normal vector through the center of the spot is $\hat{n}$ and the unit vector along the line of sight is $\hat{k}$. The angle between $\hat{n}$ and the line of sight is $\psi$, while the emission angle $\alpha$ is the angle between the photon momentum $k_0$ and the surface normal $\hat{n}$.}
\end{figure}

Figure~\ref{fig:geometry} illustrates the geometry of the problem. We set up a spherical coordinate system with its origin at the stellar center and with the $z$-axis parallel to the spin axis of the star. We use $\theta$ to denote the inclination angle between the observer's line of sight and the stellar spin axis.

We then consider a small hot spot on the neutron star surface and use ($\phi,\zeta$) to denote its azimuthal angle and polar angle. Its instantaneous position in the fixed lab frame is described by the unit vector $\hat{n}$ that points to the spot from the stellar center. The angle between $\hat{n}$ and the line of sight is denoted by $\psi$. 
We denote the unit vector along the line of sight by $\hat{k}$, so that 
\begin{equation}
\cos\psi =\hat{k}\cdot \hat{n}=\cos\theta \cos\zeta +\sin\theta \sin\zeta \cos\phi\;.
\label{eq:psi}
\end{equation} 
Throughout this paper, we will consider the azimuthal position of the spot to rotate at the neutron star spin period $P$, such that $\phi=(2\pi/P)t$.

\subsection{The Observed Flux From a Small Spot on the Neutron-Star Surface}

Following the assumptions of the S+D approximation, we describe the spacetime of the neutron star using the Schwarzschild metric but account for the Doppler effects on the energy and intensity of radiation caused by the stellar rotation. 

The observed spectral flux from a small spot at photon energy $E$ is given by~\citep{Pout1}
\begin{equation}
dF(E)=\gamma^{-1} \delta_D^4  I^\prime _{E^\prime} (\alpha ^\prime) \sqrt{1-u} \frac{d\cos\alpha }{d\cos\psi }\frac{dS \cos\alpha }{D^2}\;,
\label{eq:flux_1}
\end{equation}
where $dS$ is the area of the spot measured in the lab frame, $D$ is the distance to the observer, and $I^\prime _{E^\prime} (\alpha ^\prime)$ is the specific intensity of radiation measured in the corotating frame on the neutron-star surface at photon energy $E^\prime$, emitted with an angle $\alpha ^\prime$ with respect to the surface normal. 
The quantities $\alpha$ and $E$ are the equivalent angle and photon energy measured in the static lab frame. If the photon momentum on the surface is $\hat{k_0}$, $\alpha^\prime$ is the angle between $\hat{k_0}$ and $\hat{n}$. 

\citet{Pout2} obtained a simple analytic relation between $\cos\alpha$ and $\cos\psi$ in the Schwarzschild metric:
\begin{equation}
\cos\alpha \approx u+(1-u)\cos\psi\;.
\label{eq:lensing}
\end{equation}
In this approximation, 
\begin{equation}
    \frac{d\cos\alpha}{d\cos\psi}=1-u
\end{equation}
and, therefore, equation~(\ref{eq:flux_1}) becomes
\begin{equation}
dF(E)=\gamma^{-1}\delta_D ^4 I^\prime _{E^\prime} (\alpha ^\prime)(1-u)^{3/2}\left[u+(1-u)\cos\psi\right]\frac{dS^\prime}{D^2}\;.
\label{eq:flux_2}
\end{equation}

The Doppler factor in equation~(\ref{eq:flux_2}) is $\delta_D\equiv 1/\gamma (1-\beta \cos\xi)$, where $\cos\xi\equiv-\sin\alpha\sin\theta\sin\phi/\sin\psi$, and 
$\gamma\equiv(1-\beta ^2)^{-1/2}$ and $\beta\equiv v/c$ are the Lorentz factor and the velocity of the spot in units of the speed of light, respectively. There is an overall inverse Lorentz factor $1/\gamma$ in equation~(\ref{eq:flux_2}) that is missing in the original formula of~\citet{Pout1}. This factor arises from the different surface areas measured by an observer at infinity and by one co-moving locally with the neutron star surface~\citep{gamma}. For a spin frequency of 200~Hz, which is typical of the NICER sources, the Doppler factor changes at most from $\delta_D\sim 1.05$ to $\delta_D\sim 0.95$ within a spin period, but it enters as the fourth power in the flux. As a result, it needs to be considered when we fit models to real observational data. However, this factor depends very weakly on the model parameters $u$, $\theta$ and $\zeta$, and therefore, does not affect significantly our exploration of degeneracies in the model parameters. For computational simplicity in the following discussion, we will assume that $\delta_D=\gamma=1$ and neglect Doppler effects. In this limit, the emitted photon energy $E^\prime$ can be related to the observed photon energy $E$ by $E/E^\prime =\delta_D \sqrt{1-u}\approx \sqrt{1-u}$, and the emission angle $\alpha^\prime$ in the corotating frame is $\cos\alpha^\prime=\delta_D \cos\alpha \approx \cos\alpha$. 

Equation~(\ref{eq:flux_2}) does not account for the time delays resulting from different paths traveled by photons emitted at different phases~\citep{tdelay}. For rapidly spinning neutron stars with periods as short as $P=3$ ms, the ignored propagation time-delay effects produce a $8\%$ distortion in the light curves at large inclination and spot angles. For the slowly rotating pulsars of interest in this paper, this effect is even smaller ($\lesssim 2.5$\%; see Appendix~\ref{AppendixA}) and is subdominant compared to the other effects we will be exploring. 

For a hot spot of infinitesimal area, equation~(\ref{eq:flux_2}) can be applied only when the spot is visible by the distant observer. This condition is satisfied only when the angle between the photon direction of propagation $\hat{k}$ and the normal $\hat{n}$ on the surface is $0\le cos\alpha^\prime\le 1$, i.e., when 
\begin{equation}
\cos\alpha^\prime \simeq \cos\alpha \simeq u+(1-u)\cos\psi\ge 0\;.
\end{equation}
When the spot is not small, it is possible that only a part of it is visible at any rotational phase. In this case, one would need to use this formula to calculate the flux from each differential area on the spot based on the visibility of that region of the spot and integrate over the whole emission area numerically.

Finally, it is important to emphasize that the approximate relation~(\ref{eq:lensing}) used in the analytic formula is obtained from a Taylor expansion of the full expression around $\cos\alpha=1$. At small or negative values of $\cos\alpha$, this relation will deviate from the correct result. It is, however, adequate for the our purposes here as $\cos\alpha\approx0$ happens only when the spot is at the boundary of the visible area. This period of time occupies only a very small portion of a whole rotation period. Moreover, in a two-spot model, the flux from one of the spots, when it is at the boundary of the visible area, is usually negligible compared to the flux from the other spot, for which $\cos\alpha\approx 1$ as that spot is approaching the observer at the same time. 

In Appendix~A we compare this analytic model to a numerical solution of the full ray-tracing problem in order to assess the accuracy of our approximations (see also \citealt{Pout3}) and show that the error due to the analytic approximation is $\lesssim 2\%$ for any configuration of interest. Therefore, the use of the analytic model does not qualitatively affect any of the conclusions presented in this paper. 

\subsection{The Radiation Emerging From the Stellar Surface}

The specific intensity $I^\prime _{E^\prime} (\alpha ^\prime)$ of radiation emerging from the neutron star surface at energy $E^\prime$ and at angle $\alpha^\prime$ with respect to the surface normal depends on the temperature profile throughout the atmosphere. We assume here a blackbody spectrum of radiation at an effective temperature $T^\prime$, as measured locally, with a beaming function that is, in general, non-isotropic. 

\begin{figure}[t]
	\centering
	\includegraphics[width=1\linewidth, keepaspectratio]{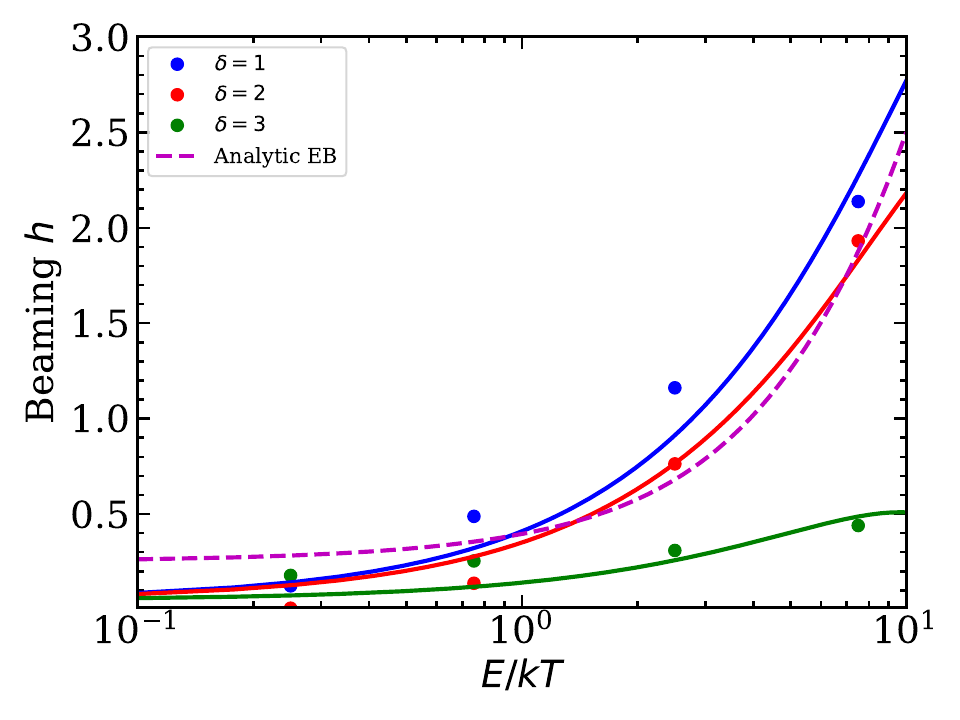}
	\caption{\label{fig:beaming} The beaming factor $h$ as a function of photon energy (in units of the effective temperature of the atmosphere), for neutron-star atmospheres bombarded by relativistic leptons with a power-law energy spectrum of index $\delta$. The filled circles are the results of fitting eq.~(\ref{eq:beaming}) to the calculations of~\citet{atmosphere}, while the solid curves are quadratic fits to these data points.  The dashed curve shows the approximate scaling of the beaming factor with photon energy for a deep-heated Eddington-Barbier atmosphere that we derive in Appendix~B.}
 \label{fig:beaming}
\end{figure}

We introduce a beaming factor $h(E^\prime,T^\prime)$ that depends on both temperature and photon energy to specify the degree of anisotropy, so that the specific intensity becomes
\begin{equation}
I^\prime _{E^\prime} (\alpha ^\prime)=I_b(E^\prime,T^\prime)\frac{1+h(E^\prime,T^\prime)\cos\alpha^\prime}{1+\frac{2}{3}h(E^\prime,T^\prime)}\;,
\label{eq:beaming}
\end{equation}
where 
\begin{equation}
I_b(E^\prime,T^\prime)\equiv\frac{2E^{\prime 3}}{h^3c^2(e^{E^\prime/kT^\prime}-1)}
\end{equation}
is the specific intensity of blackbody radiation and $k$ and $h$ are the Planck and Boltzmann constants, respectively. The denominator $1+2/3h(E^\prime,T^\prime)$ is used in this expression to preserve the total flux emerging from the surface,
\begin{equation}
    F(E^\prime,T^\prime)=\pi\int_{\alpha^\prime} I^\prime_{E^\prime} (\alpha ^\prime) \cos \alpha^\prime d\alpha^\prime=\pi I_b(E^\prime,T^\prime)\;,
\end{equation}
independent of the beaming function.

Early models of pulse profiles from rotation-powered pulsars used the spectrum and beaming of an atmosphere in radiative equilibrium that is heated from below~\citep{Miller2019,Miller2021,Riley2019,Riley2021}. However, \citet{Baubock} showed that, for polar caps heated by return magnetospheric currents, the heating can be shallower, leading to temperature inversions,  and result in a beaming function that is significantly flatter than that of atmospheres with deep heating. In such situations, the beaming function is determined primarily by the energy spectrum and low-energy cutoff of the particles in the return currents from the neutron star magnetosphere. Models of atmospheres in radiative equilibrium were then constructed by \citet{atmosphere} for heating induced by bombardment by leptons with power-law energy spectra. In Appendix~B, we devise a quadratic functional form of the beaming factor $h$ that approximates the results of the models, which is given by,
\begin{equation}
h(E^\prime,T^\prime)=a+b\left(\frac{E^\prime}{kT^\prime}\right)+c\left(\frac{E^\prime}{kT^\prime}\right)^2\;,
\label{eq:beaming_energy}
\end{equation}
with coefficients $a$, $b$ and $c$ that depend on the power law index $\delta$ of the energy distribution of particles in the return current (see Fig.~\ref{fig:beaming}). For $\delta=1$, the large number of leptons at high energies heat primarily the deep layers of the atmosphere and introduce a significant degree of beaming towards the normal ($h\gtrsim 1$). For steeper spectra of the leptons (i.e., for $\delta=3$), the distribution shifts towards lower energy particles that heat primarily the shallower layers of the atmosphere and generate a more isotropic beaming in the emerging radiation ($h\simeq 0$).

\begin{table}[t]
\caption{Physical Parameters for Neutron-Star Profile Modeling}
\begin{center}
\begin{tabular}{c c c c}
\toprule
Parameter & Description & Units & Range \\
\midrule
$M$ & Neutron-star mass & $M_\odot$ & $\sim 1-2.5$ \\ 
$R$ & Neutron-star radius & km & $\sim 10-15$ \\ 
$\theta$ & observer inclination angle & degrees & $0-90$ \\
$\zeta$ & spot colatitude & degrees & $0-180$ \\
$T$ & effective temperature & keV & $>0$\\
$dS$ & spot area & m$^2$ & $>0$ \\ 
$D$ & distance & kpc & $>0$ \\ 
\bottomrule
\end{tabular}
\end{center}
\label{table:1}
\end{table}

\section{Reducing the Degeneracy of Model Parameters}

As discussed in the previous section, the flux observed at a given photon energy $E$ and rotational phase $\phi$ for the symmetric case of antipodal hot spots with equal sizes and temperatures on a neutron star depends on seven model parameters, which are summarized in Table~\ref{table:1}. In practice, however, there exist several degeneracies between these model parameters that reduce the number of independent quantities that uniquely determine the pulse profiles and, thus, that can be measured from the pulse profiles. (Note that even though we use a simple configuration to establish the origin of the degeneracies, their presence is exacerbated if we introduce additional complexity to the configuration, such as non-antipodal hot spots with different sizes and temperatures.)

We now use the analytic model to identify the parameter degeneracies. We achieve this by looking at the Fourier content of the pulse profiles, which is an approach that has been very fruitful in the past~\citep{Pout1,Psaltis2014,Feryal1}. 

\begin{table}[t]
\caption{Weakly Degenerate Parameters}
\begin{center}
\begin{tabular}{c c c c}
\toprule
Parameter & Description & Units & Range \\
\midrule
$q$ & $u+(1-u)\cos\theta \cos\zeta$ & dimensionless & $-1<q<1$ \\
$s$ & $u-(1-u)\cos\theta \cos\zeta$ & dimensionless & $-1<q<1$ \\
$p$ & $(1-u)\sin\theta \sin\zeta$ & dimensionless & $0<p<1$ \\ 
$T_\infty$ & $T\sqrt{1-u}$ & keV & $>0$ \\
$A$ & $dS/D^2$ & dimensionless & $>0$\\
\bottomrule
\end{tabular}
\end{center}
\label{table:2}
\end{table}

For a slowly spinning neutron star, there are two obvious degeneracies. First, the external spacetime depends only on the compactness $u=2 GM/Rc^2$ and not on the mass or radius individually. Similarly, it is straightforward to see from equation~(\ref{eq:flux_2}) that the distance $D$ to the source and the surface area of the hot spot $dS$ cannot be distinguished with respect to their effects on the predicted pulse profiles. For this reason, we combine the mass and radius in the compactness $u$ and the distance and surface area into one parameter $A\equiv dS/D^2$, which measures the solid angle subtended by a surface area equal to that of the hot spot at the distance of the source. 

Besides these trivial degeneracies, there are others that require a more detailed analysis. For example, the compactness by itself is not adequate to remove the degeneracies that are inherent in the gravitational lensing effects, which also depend on the geometry of the configuration. As equation~(\ref{eq:flux_2}) shows, gravitational lensing determines both the projected surface area of the hot spot at a given rotational phase through the term $dS \cos\alpha$ and the magnitude of the intensity of radiation seen from the neutron star surface through the beaming function $I^\prime_{E^\prime}(\alpha^\prime)$. However, both of these effects depend on the cosine of the angle $\alpha$ (or, equivalently, $\alpha^\prime$), which we can use equations~(\ref{eq:lensing}) and (\ref{eq:psi}) to write as
\begin{equation}
    \cos\alpha=q+p \cos\phi\;,
\end{equation}
where
\begin{eqnarray}
    q&\equiv& u+(1-u) \cos\theta\cos\zeta\;,\\
    p&\equiv& (1-u) \sin\theta\sin\zeta\;.
\end{eqnarray}
The quantity $q$ measures the influence of gravitational lensing on the typical projection of the hot-spot surface area as viewed by the distant observer, i.e., $\langle \cos\alpha^\prime dS\rangle=q dS$ (with the last equality being formally valid when the hot spot is visible throughout the neutron-star rotation). The quantity $p$ measures the amplitude of the oscillation in the projection of the hot-spot surface area as a function of rotational phase.

For a configuration with two antipodal hot spots, the polar angle of the ``far'' hot spot is equal to $\zeta_2=\pi-\zeta$, such that $\sin\zeta_2=\sin\zeta$ but $\cos\zeta_2=-\cos\zeta$. As a result, the typical angle of projection of the ``far'' hot spot is measured by the physical quantity
\begin{equation}
    s\equiv u+(1-u) \cos\theta\cos\zeta_2 = u-(1-u) \cos\theta\cos\zeta\;,
\end{equation}
whereas the amplitude of oscillation in the angle of its projection remains equal to $p$.

Finally, because both the intensity of blackbody radiation as well as the beaming factor are primarily functions of the quantity $(1-u)^{-1/2}(E/T^\prime)$, it is convenient to define a temperature at infinity as $T_\infty=T^\prime(1-u)^{1/2}$ and use this as an appropriate variable. 

These simple arguments reduced the number of independent parameters of the model from 7 (as shown in Table~\ref{table:1}) to 5, which are summarized in Table~\ref{table:2}. We now proceed to providing expressions for the flux observed from a neutron star with two antipodal spots as a function of photon energy $E$ and rotational phase angle $\phi$ using only the combinations of model parameters shown in the latter table. In the following sections, we use Bayesian parameter estimation to fit synthetic NICER data to demonstrate how these parameters indeed reduce degeneracies and allow us to explore potential biases and systematic uncertainties in the inferences of neutron-star properties.

\subsection{The Fourier Content of Pulse Profiles}

The surface thermal emission is thought to arise from localized ``hot spots'' within a sub-region of each of the polar caps. For a pure dipole field configuration, the polar caps are circles centered on the magnetic poles when the star is slowly rotating \citep{Arons2013}. For a more complex multipolar field structure, the hot spots may become ring-like or have irregular shapes~\citep{Will}. Analysis of pulse profiles observed by NICER suggest such complex magnetic field structures~\citep{Miller2019,Riley2019}. 

For the purposes of our exploration, we focus on the simple geometry with two antipodal hot spots of the same size and effective temperature. Under these assumptions, the colatitude of the second spot is $\pi-\zeta$ and its longitude is $\pi +\phi$. The total flux observed at infinity is
\begin{equation}
F(E,\phi)=F_1(E,\phi)+F_2(E,\phi),
\end{equation}
where $F_1$ and $F_2$ are the fluxes from the first and second spot, respectively. 

Using equation~(\ref{eq:flux_2}), we write for the flux observed from the first spot
\begin{equation}
F_1(E,\phi)=
\begin{cases}
    {\cal F}_1(E,\phi)\;, &\text{if } u+(1-u)\cos\psi>0\\
    0\;, & \text{otherwise}
    \label{eq:F_1}
\end{cases}
\end{equation}
where
\begin{equation}
{\cal F}_1(E,\phi)=q(1+\bar{h}q)\bar{F}(E)\left[1+r_1(E)\cos\phi+r_2(E)\cos^2\phi\right]
\label{eq:Fbar}
\end{equation}
\begin{equation}
\bar{F}(E)\equiv \frac{A(1-u)^{3/2}}{1+(2/3)\bar{h}} I^\prime _b(E^\prime ,T),
\end{equation}
\begin{equation}
r_1(E)\equiv \frac{p}{q}\left(1+\frac{\bar{h}q}{1+\bar{h}q}\right)\;,
\label{eq:r1}
\end{equation}
\begin{equation}
r_2(E)\equiv \left(\frac{p}{q}\right)^2\frac{\bar{h}q}{1+\bar{h}q}\;,
\label{eq:r2}
\end{equation}
and the beaming factor is evaluated at
\begin{equation}
    \bar{h}\equiv h(E/\sqrt{1-u},T).
\end{equation}

Similarly, the flux $F_2(E)$ from the second spot can be derived from the same equation by using the antipodal symmetry ($q\rightarrow s, \phi \rightarrow \phi+\pi$): 
\begin{equation}
F_2(E,\phi)=
\begin{cases}
    {\cal F}_2(E,\phi)\;, &\text{if } u-(1-u)\cos\psi>0\\
    0\;, & \text{otherwise}
    \label{eq:F_2}
\end{cases}
\end{equation}
where
\begin{equation}
{\cal F}_2(E,\phi)=s(1+\bar{h}s)\bar{F}(E)\left[1-t_1(E)\cos\phi+t_2(E)\cos\phi^2\right]
\end{equation}
\begin{equation}
t_1(E)=\frac{p}{s}\left(1+\frac{\bar{h}s}{1+\bar{h}s}\right),
\label{eq:t1}
\end{equation}
\begin{equation}
t_2(E)=\left(\frac{p}{s}\right)^2\frac{\bar{h}s}{1+\bar{h}s}\;.
\label{eq:t2}
\end{equation}

The two conditions in the expressions from the fluxes observed from the two spots account for their window of visibility to the distant observer. To evaluate these conditions explicitly in terms of the model parameters, we assume, without loss of generality, that the first spot is visible at zero phase and that the observer is in the north hemisphere so that $\theta <90^{\circ}$ and $\zeta <90^{\circ}$. Following~\citet{Pout1}, we identify four visibility classes: {\em (I)\/} Only the first spot is visible at all  times while the second spot is not visible, i.e., $F_1>0$ but $F_2=0$; {\em (II)\/} The first spot is always visible, i.e., $F_1>0$, but the second spot is visible only during phases $\phi _2<\phi<2\pi-\phi_2$; {\em (III)\/} Both spots are always visible, i.e., $F_1>0$ and $F_2>0$; {\em (IV)\/} Neither of the two spots is always visible, with the first spot disappearing at a phase angle $\phi_1$.

We can use the visibility condition in equation~(\ref{eq:F_2}) to evaluate the value of $\phi_2$ at which the second spot becomes visible as 
\begin{equation}
\cos\phi_2=\max\left[\frac{s}{p},-1\right]
\label{eq:phi2}
\end{equation}
and equation~(\ref{eq:F_1}) to evaluate the value of $\phi_1$ at which the first spot disappears as
\begin{equation}
    \cos\phi_1=\max\left[\frac{-q}{p},-1\right]\;.
    \label{eq:phi1}
\end{equation}
Using these expressions, we can write the flux observed at infinity as
\begin{equation}
F_{total}(E,\phi)=
     \begin{cases}
       {\cal F}_1(E,\phi), &\text{if } \phi<\phi_2 \\&\text{or } \phi>(2\pi-\phi_2)\\
       {\cal F}_1(E,\phi)+{\cal F}_2(E,\phi),  &\text{if } \phi_2\le\phi\le\phi_1 \\&\text{or } (2\pi-\phi_1)\le\phi\le(2\pi-\phi_2)\\
       {\cal F}_2(E,\phi),  &\text{if } \phi_1<\phi<(2\pi-\phi_1),
     \end{cases}
\end{equation}
which encompasses all visibility classes.

Finally, we perform an expansion of the flux in terms of its Fourier components as 
\begin{equation}
F(E,\phi)=A_0(E)+\sum_{n=1} A_n(E)\cos\left(n \phi\right)
\end{equation}
where we have suppressed the expansion into sines, since we have neglected the Doppler effects, which are the ones contributing to the sine expansion~\citep[see][]{Pout1}. The explicit expressions for the first few harmonics are
\begin{equation}
\begin{split}
A_0(E)=&\frac{q(1+\bar{h}q)\bar{F}(E)}{2\pi}\times\\\
&\qquad \left[(2+r_2)\phi_1+(2r_1+r_2\cos\phi_1)\sin\phi_1\right]\\
&+\frac{s(1+\bar{h}s)\bar{F}(E)}{2\pi}\times\\
 &\qquad\left[(2+t_2)(\pi-\phi_2)+(2t_1-t_2\cos\phi_2)\sin\phi_2\right]\;,
\end{split}
\label{eq:A0}
\end{equation}
\begin{equation}
\begin{split}
A_1(E)=&\frac{q(1+\bar{h}q)\bar{F}(E)}{\pi}\times\left[r_1\phi_1\right.\\
&\left.\qquad+\left(2+r_1\cos\phi_1+\frac{3}{2}r_2\right)\sin\phi_1+\frac{1}{6}r_2\sin(3\phi_1)\right]\\
&+\frac{s(1+\bar{h}s)\bar{F}(E)}{\pi}[t_1(\phi_2-\pi)\\
& \left.\qquad+\left(-2+t_1\cos\phi_2-\frac{3}{2}t_2\right)\sin\phi_2-\frac{1}{6}t_2\sin(3\phi_2)\right]\\
\end{split}
\label{eq:A1}
\end{equation}
and
\begin{equation}
\begin{split}
A_2(E)=&\frac{q(1+\bar{h}q)\bar{F}(E)}{\pi}\left[\frac{1}{2}r_2\phi_1+r_1\sin(\phi_1)\right.\\
&\qquad+\left(1+\frac{1}{2}r_2\right)\sin(2\phi_1)\\
&\qquad+\left.\frac{1}{3}r_1\sin(3\phi_1)+\frac{1}{8}r_2\sin(4\phi_1)\right]\\
&+\frac{s(1+\bar{h}s)\bar{F}(E)}{\pi}\left[\frac{1}{2}t_2(\phi_2-\pi)+t_2\sin(\phi_2)\right.\\
& \qquad +\left(-1+\frac{1}{2}t_2\right)\sin(2\phi_2)\\
&\qquad+\left.\frac{1}{3}t_1\sin(3\phi_2)+\frac{1}{8}t_2\sin(4\phi_2)\right]\;.
\end{split}
\label{eq:A2}
\end{equation}
For visibility Class~I, for which $\phi_2\rightarrow \pi$ and $\phi_1\rightarrow \pi$, these reduce to 
\begin{eqnarray}
    A_0(E)&=&q(1+\bar{h}q)\bar{F}(E)\left(1+\frac{r_2(E)}{2}\right)\\
    A_1(E)&=&q(1+\bar{h}q)\bar{F}(E)r_1(E)\\
    A_2(E)&=&\frac{1}{2}q(1+\bar{h}q)\bar{F}(E)r_2(E)\;,
\end{eqnarray}
which provide a useful scale for the relative magnitudes of the various Fourier components.

\begin{figure*}[t]
	\centering
	\includegraphics[width=0.8\textwidth, keepaspectratio]{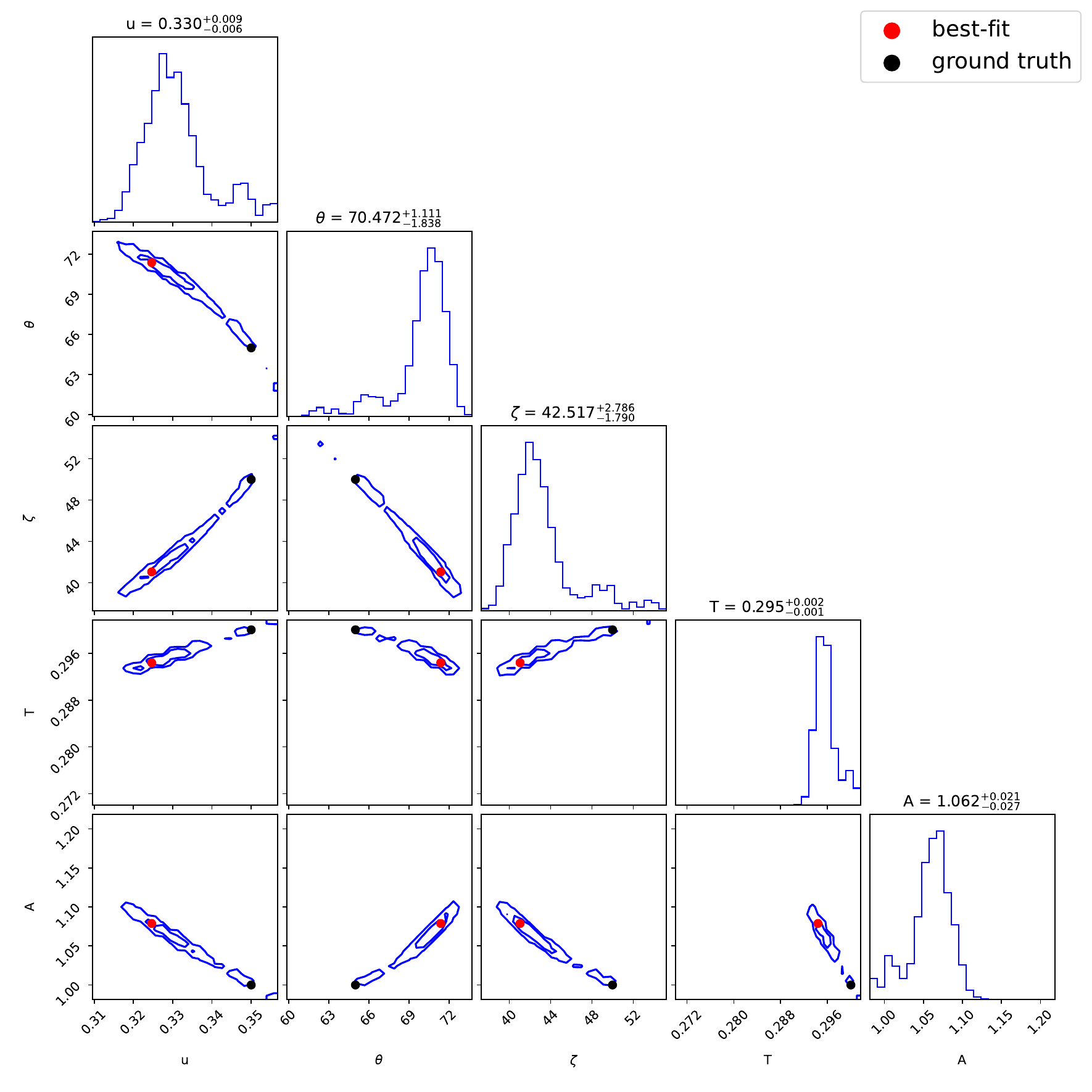}
	\caption{\label{fig:6550zb0} The posterior of fitting synthetic pulse profile data in the energy range $0.3\;$keV$-1.5\;$keV, using a model described in  terms of the physical parameters shown in Table~\ref{table:1}. In each panel, the contours represent the 68-th and 95-th percentiles of the marginalized posterior distributions. The synthetic data were generated for isotropic beaming ($h=0$) and $u=0.35$, $T=0.3$~keV, $A=1$, $\theta=65^\circ$ and $\zeta=50^\circ$, as shown with the black dots on the plots. The best-fit values are shown with red dots. There is significant degeneracy between the various model parameters, introducing substantial biases in the marginalized distributions.}
\end{figure*}

\section{Synthetic Data and Bayesian Parameter Estimation}

In order to explore the influence of different model parameters and choices on the inference of neutron-star properties via pulse profile modeling, we generate various sets of synthetic data based on the two-antipodal-spot model. When choosing the values for the various parameters, we use the recent fit results to NICER data as a guide~\citep{Miller2019,Miller2021,Riley2019,Riley2021}. 

To generate realistic synthetic data, we include corrections due to the interstellar extinction as well as a realistic response matrix for NICER\footnote{Specifically, we use \texttt{nixtiaveonaxis20170601v002.arf} and  \texttt{nixitref20170601v001.rmf}}. We calculate the interstellar extinction based on an early, simple model~\citep{Robert1983} and fix the column density of hydrogen to $N_{\rm H}=1.5\times 10^{20}\;$cm$^{-2}$, which is close to the fitted value for PSR~J0740+6620 \citep{Miller2021,Riley2021}. For computational efficiency, we bin the energy channels into 24 energy bins of $0.05\;$keV width in the NICER energy band between $0.3\;$keV and $1.5\;$keV. Finally, we calculate synthetic data in 32 phase intervals per cycle. 

The number of photons in each energy and time bin, before applying the instrument response, is given by $N=F(E,\phi)\Delta t \Delta E$, where $F(E,\phi)$ is the flux, and $\Delta t$ and $\Delta E$ are the widths of each time bin and energy bin, respectively. In order to generate synthetic data with realistic statistical errors, we draw a value for each bin from a Gaussian distribution centered at the ground-truth photon number and with a standard deviation equal to $\sqrt{N}$. 

\begin{figure*}[t]
	\centering
	\includegraphics[width=0.8\textwidth, keepaspectratio]{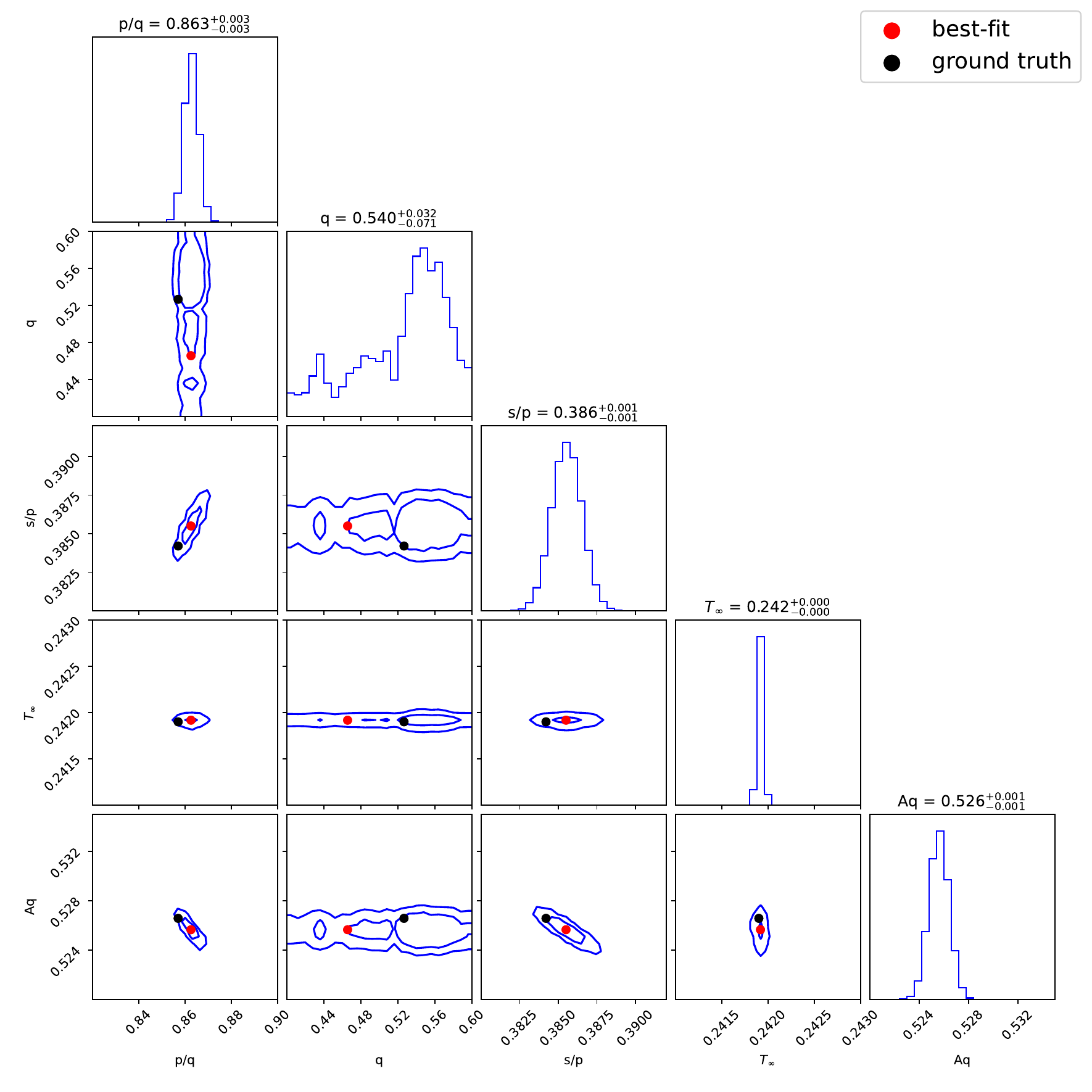}
	\caption{\label{fig:6550new} The posterior of fitting the same synthetic data as those in Fig.~\ref{fig:6550zb0} but for a model described in terms of the weakly correlated parameters shown in Table~\ref{table:2}. The degeneracies between four of the model parameters have largely disappeared, while one parameter ($q$) is practically unconstrained. The synthetic data correspond to $p/q=0.857$, $q=0.523$, $s/p=0.384$, $T_\infty=0.24$ and $Aq=0.527$. Here, the overall normalization factor $A$ is scaled by the value in our fiducial model.}
\end{figure*}

For our fiducial model, we set the distance to the source to $d=1.2 \; \text{kpc}$ and the spot area to $dS=4.52\times 10^6 \; \text{m}^2$~\citep{Miller2021,Riley2021}. We also set the total exposure time to $t=$1602683\;s, i.e., the same as the observations. For a blackbody temperature of $T=0.15$~keV, a mass of $M=2.07~M_\odot$ and a radius of $R=11.5~$km, the total number of photons in the synthetic data is equal to 310000.

As we vary the model parameters to explore their effects on potential biases, the total number of photons, and hence the statistical errors, will also vary and depend on the particular values of the parameters. In order to partially decouple the effect of varying the model parameters from that of altering the measurement uncertainties, we adjust the overall normalization of the spectrum to obtain, for each configuration, a total number of photons that is comparable to that of the fiducial spectrum. 

For each set of synthetic data, we perform Bayesian parameter inference using a Metropolis-Hastings Markov-Chain Monte Carlo (MCMC) algorithm with steps drawn from Gaussian distributions and no tempering. Our algorithm is loosely based on \texttt{MARCH}; \citealt{Psaltis2022}. For the priors over the model parameters, we use flat-top distributions with limits shown in Tables~\ref{table:1} or ~\ref{table:2}, as appropriate. 

\begin{figure*}[t]
	\centering
	\includegraphics[width=0.8\textwidth, keepaspectratio]{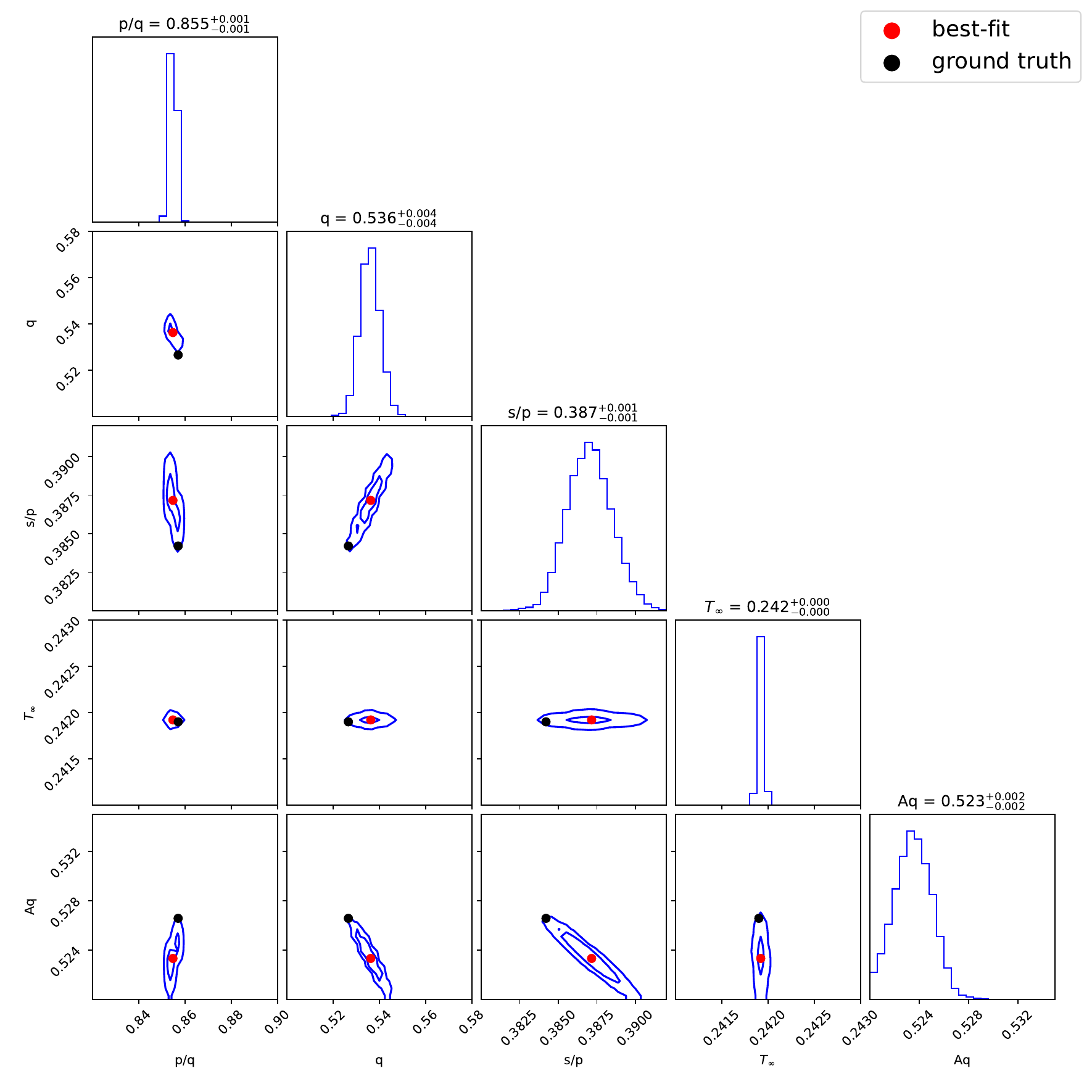}
	\caption{\label{fig:6550h} The posterior of fitting synthetic data with the same neutron-star parameters as in Fig.~\ref{fig:6550new} but for non-isotropic, energy independent beaming with $h=1.0$. The presence of the non-isotropic beaming allows for the parameter $q$ to be constrained and breaks the degeneracy between the model parameters.}
\end{figure*}

\section{From Observables to Neutron-Star Parameters}

Our ultimate goal is to use observations of pulse profiles to constrain the properties of neutron stars. In the previous section, we showed that five parameters can, in principle, be inferred from observations within the dipolar hot spot model. However, the pulse profiles do not show the same degree of sensitivity to each of these parameters. Therefore, the parameters cannot be inferred with the same degree of confidence. Furthermore, this sensitivity changes in response to the properties of the neutron star, such as the beaming of radiation emitted from the polar caps, or to the characteristics of the observations, such as the observed range of photon energies. In this section, we identify the salient features in the observations that allow us to constrain the individual model parameters and discuss the resulting impact on parameter inference. 

\subsection{Degeneracies between Parameters}
\label{sec:beaming}

To illustrate the large correlations between the physical model parameters (Table~\ref{table:1}) and the usefulness of the weakly degenerate parameters introduced in Table~\ref{table:2}, we first consider the case of isotropic beaming. This is the beaming function assumed in early work and exacerbates the unforeseen degeneracies, as we will show.  

For this configuration, we generate synthetic data for a neutron star with compactness $u=0.35$, atmospheric temperature $T=0.3$~keV, observer inclination $\theta=65^\circ$, and colatitude of the hot spot $\zeta=50^\circ$. Figure~\ref{fig:6550zb0} shows the corner plot for the posteriors we obtain when we fit the analytic model to this synthetic dataset. Here, the overall normalization factor $A$ is scaled by the value in our fiducial model such that $A=1$ corresponds to a value of $dS/D^2$ that is the same as in the fiducial model. The contours of the various cross sections of the posteriors expand along narrow curves containing both the best-fit and the ground-truth values, signifying a further degeneracy between model parameters, which are beyond the trivial degeneracies contained in the definitions of the parameters $u$ and $A$. 

This result is easy to understand using the analytic formulae~(\ref{eq:F_1})-(\ref{eq:t2}). We see that, in the case of isotropic beaming, i.e., when $\bar{h}=0$, the three parameters $q$, $s$, and $p$ that describe gravitational lensing and the overall normalization $A$ appear only in three combinations $p/q$ (eqs.~[\ref{eq:r1}], [\ref{eq:r2}], [\ref{eq:phi1}]), $p/s$ (eqs.~[\ref{eq:t1}], [\ref{eq:t2}], [\ref{eq:phi2}]), and $qA$ (eqs.~[\ref{eq:F_1}]-[\ref{eq:Fbar}]). In other words, the observed profiles carry enough information to constrain only three out of these four parameters (or their combinations). 

We demonstrate this degeneracy explicitly in Fig.~\ref{fig:6550new}, where we show the posteriors of the same Bayesian inference but displayed in terms of the parameters $q$, $p/q$, $s/p$, $T_\infty$, $A\;,q$. First, this representation of the posteriors clearly shows that only weak correlations remain between the newly introduced model parameters, which enables very efficient sampling with an MCMC method. Second, it demonstrates that the values of only four out of the five parameters can be tightly bound by the data but the fifth one ($q$ in this case) is unconstrained. 

In eight out of the ten 2-dimensional marginalized posteriors shown in the corner plot of Figure~\ref{fig:6550new}, the ground truth parameters lie within or on the 65-th percentile contours, as expected. The differences between inferred and ground-truth values are the result of the particular representations of the formal measurement errors. Because of these differences, however, and the complete degeneracy between the physical model parameters shown in Figure~\ref{fig:6550zb0}, the marginalized posterior over the neutron-star compactness $u$ has a width of $(+0.009,-0.006)$ but is biased by $-0.02$ towards lower values (i.e., towards larger radii than the ground truth). If the ground truth model corresponded to a radius of 11~km and the neutron-star mass was known a priori, the formal error in the radius measurement would be $\sigma_{R}\sim (+0.009,-0.006)/0.35\times 11$~km$\simeq (+0.28,-0.18)$~km, whereas the bias would have been equal to $\delta R\sim 0.02/0.35\times 11$~km$\simeq 0.6$~km, i.e., three times larger.

\begin{figure}[t]
	\centering
	\includegraphics[width=1\linewidth, keepaspectratio]{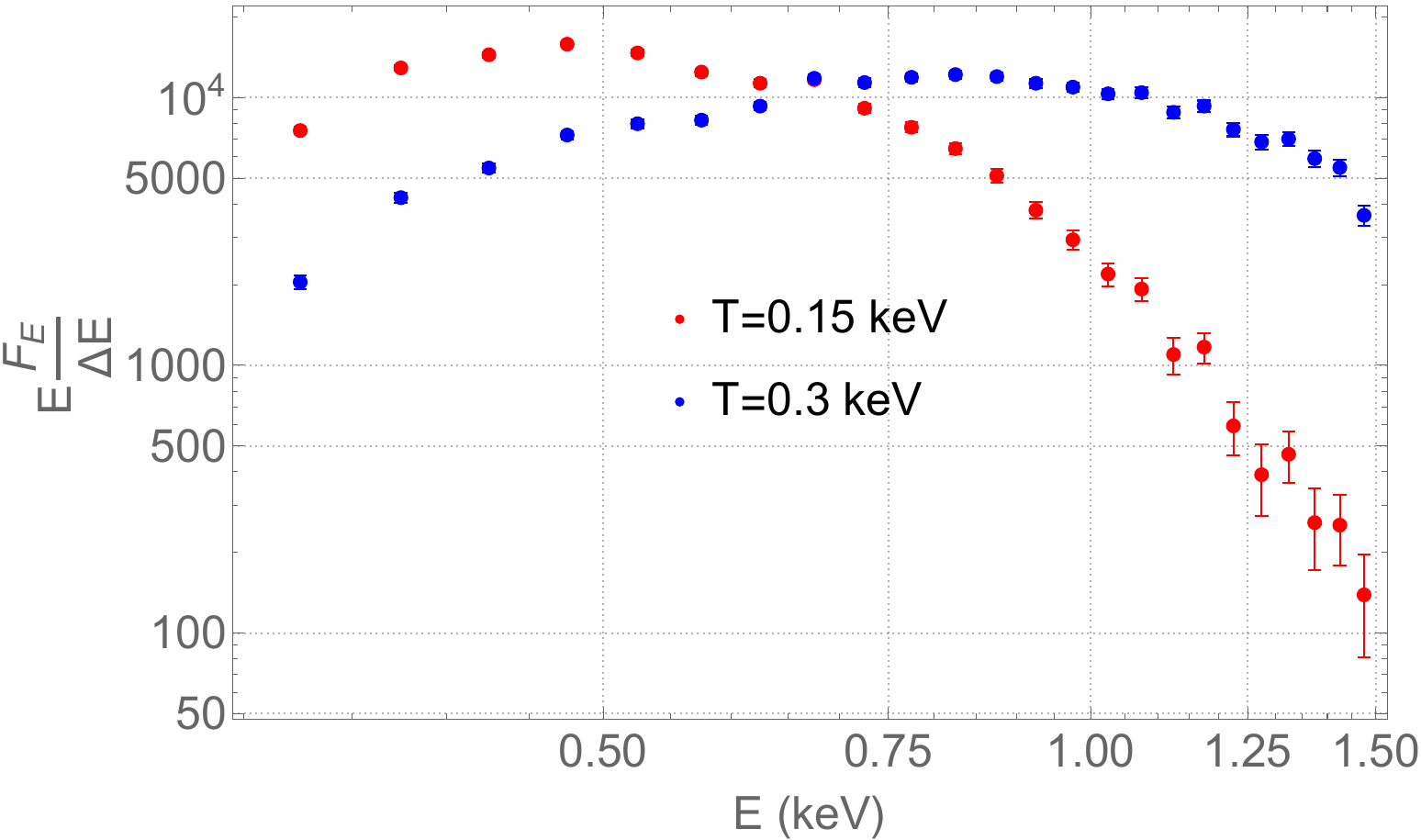}
	\caption{\label{fig:spectrum} The time-averaged, synthetic NICER countrate spectrum for two neutron stars with different temperatures, normalized such that the total number of photons in each spectrum is the same. At $T=0.15$~keV, the rapid drop of the spectrum at the higher energies increases substantially the measurement uncertainties.}
\end{figure}

\subsection{The Importance of Energy-Dependent Beaming}

The complete degeneracy between the parameters shown in the previous example can only be broken if the radiation emerging from the neutron star is non-isotropic. Non-isotropic beaming provides additional information that allows constraints on each of the model parameters, although the constraints remain weakly correlated. 

\begin{figure}[t]
	\centering
	\includegraphics[width=1\linewidth, keepaspectratio]{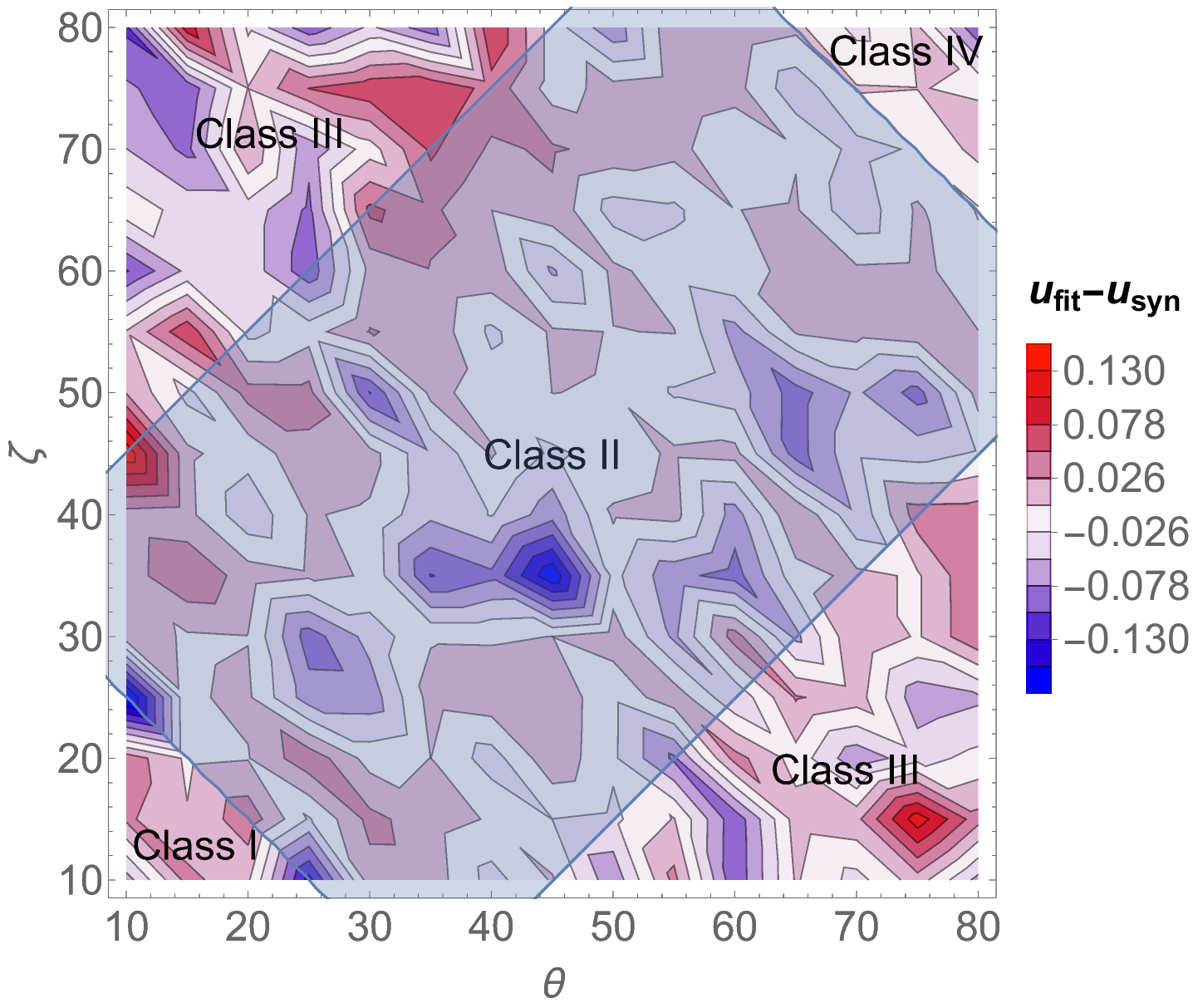}
	\caption{\label{fig:bias0} Contours of the difference between the fitted compactness $u=2GM/Rc^2$ and its ground-truth value for a configuration with $u_{\rm syn}=0.45$, $T=0.15$~keV, and different observer inclinations ($\theta$) and colatitudes of the polar caps ($\zeta)$. The $\theta -\zeta$ plane is divided into different parts corresponding to different classes of visibility. The difference between the fitted and ground-truth values of the compactness does not depend on the geometric parameters of the model nor on the visibility class of the configuration.}
\end{figure}

We illustrate this in Figure~\ref{fig:6550h}, where we show the posteriors of fitting synthetic data from a neutron star with the same parameters as in the previous examples but with energy-independent, non-isotropic beaming with $h=1.0$. In this case, there is significant information in the pulse profiles to allow for the parameter $q$ to be partially constrained, therefore breaking the complete degeneracy between the various model parameters (see, though, \S\ref{sec:energyrange}). However, this example also demonstrates the impact of the beaming function on the magnitude of the uncertainties in the inferred model parameters and the importance for parameter inference of an {\em a priori\/} theoretical knowledge of the radiation beaming. We explore this aspect of pulse profile modeling in detail in a separate publication~\citep{Zhao2024}

\subsection{The Impact of The Observed Energy Range}
\label{sec:energyrange}

In the example we discussed in \S\ref{sec:beaming}, we set the temperature of the blackbody to $T=0.3$~keV. This is similar to the inferred temperature of PSR~J0437$-$4715~\citep{thermX3}, which has been the prototypical source for inferring neutron-star parameters via fitting pulse profiles, but is higher by a factor of $\sim 3$ compared to the  temperatures inferred for  PSR~J0030$+$0451~\citep{Riley2019,Miller2019} and PSR~J0740+6620~\citep{Riley2021,Miller2021}, for which NICER measurements have been initially reported. This difference is significant because the peak of the blackbody photon spectrum for the first source falls within the observed range of photon energies for NICER, allowing for the temperature ($T_\infty$) and normalization ($A$) of the spectrum to be well constrained. This is not the case for the last two sources, for which the observed energy range is only sensitive to the Wien tail of the spectrum.

\begin{figure}[t]
	\centering
	\includegraphics[width=1\linewidth, keepaspectratio]{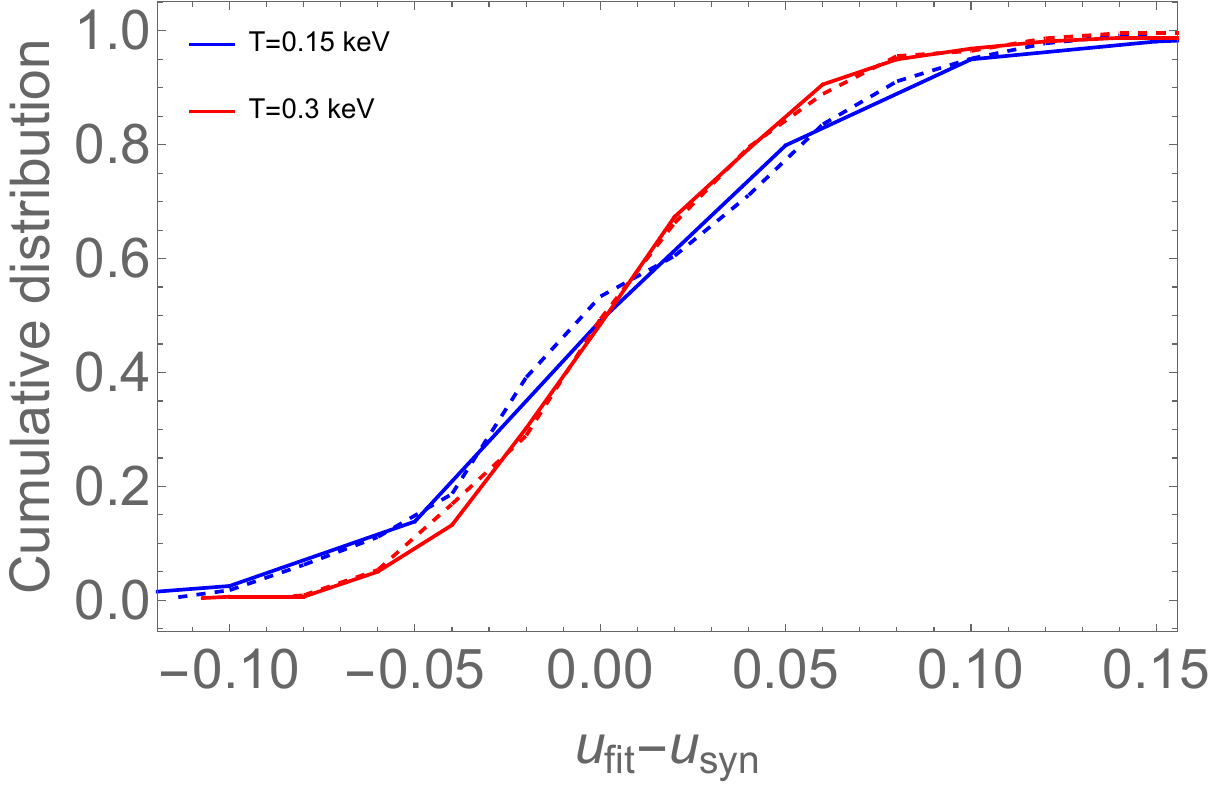}
	\caption{\label{fig:bias1} Cumulative histograms of the differences $u_{\rm fit}-u_{\rm syn}$ between the fitted compactness $u_{\rm fit}=2GM/Rc^2$ and its ground-truth value for two configurations with $u_{\rm syn}=0.45$. The blue curves corresponds to a cooler neutron star ($T=0.15$~keV) while the red lines curves to a hotter neutron star ($T=0.30$~keV). Solid curves were calculated by sampling the range of geometric parameters shown in Fig.~\ref{fig:bias0}, whereas dashed curves correspond to 225 realizations of noise for $(\theta,\zeta)=(65^\circ,50^\circ)$. The total number of photons has been kept the same for all configurations. The assumed energy range of observations is $(0.3-1.5)$~keV.  Utilizing an energy range that encompasses the peak of the neutron-star spectrum, as is the case of the $T=0.3$~keV configuration, significantly decreases systematic errors in the inference of its compactness.}
\end{figure}

To explore the effect of the energy range of the observation, we make two sets of synthetic data based on the two-antipodal-spot model in the energy range $0.3\;$keV to $1.5\;$keV, for stars with compactness $u=0.45$, a realistic energy-dependent beaming two different effective temperatures $T=0.15$~keV and $T=0.30$~keV. (We set the first effective temperature to 0.15~keV instead of 0.1~keV to account for the color-correction factor, because we assume a blackbody emission rather than using a hydrogen atmosphere emission model). 

Figure~\ref{fig:spectrum} compares the time-averaged synthetic NICER spectra for the two blackbody temperatures, when we set the observer inclination to $\theta=65^\circ$ and the colatitude of the spots to $\zeta=50^\circ$ but keep the total number of photons the same. The rapid decrease of the low-temperature spectrum at energies $\gtrsim 1.0$~keV substantially increases the formal errors in these energy bins, from $\sim 1\%$ for the higher temperature spectrum to $\sim 8\%$ for the lower temperature one, degrading the quality of the measurements and, as we will now show, of the constraints on the model parameters. 

To quantify the influence of the observed energy range on the systematics introduced on the inferred model parameters, we simulated synthetic data for a broad range of the geometric model parameters (i.e., for $10^\circ\le \theta\le 80^\circ$ and $10^\circ\le \theta\le 80^\circ$) while keeping the same total number of observed photons. We then performed MCMC analysis of each data set using the analytic model and calculated the difference between the compactness $u_{\rm fit}$ with the highest posterior and the ground-truth value used in the synthetic data $u_{\rm syn}$. Figure~\ref{fig:bias0} shows contours of this difference on the $\theta-\zeta$ plane for the neutron star with $T=0.15$~keV, demonstrating that the inferred compactness can be larger or smaller than the ground truth value by as much as $\vert u_{\rm fit}-u_{\rm syn}\vert\sim 0.1$, with no apparent dependence on the geometry of the configuration or its visibility class. This systematic uncertainty is larger by a factor of a few than the formal uncertainties in the inferred values of the compactness.

Figure~\ref{fig:bias1} shows (solid curves) the cumulative distributions of the values for the difference $u_{\rm fit}-u_{\rm syn}$ obtained by changing the geometric model parameters for both the cooler ($T=0.15$~keV) and the hotter ($T=0.30$~keV) neutron stars. The dashed curves in the same figure show the cumulative distributions of the same quantities but for 225 realizations of noise for a single set of geometric model parameters, demonstrating again the fact that this systematic uncertainty does not depend on the geometry of the system. The two sets of distributions shown in solid lines correspond to $u_{\rm fit}-u_{\rm syn}=0.003^{+0.055}_{-0.050}$ for the cooler star and to $u_{\rm fit}-u_{\rm syn}=-0.002^{+0.013}_{-0.031}$ for the hotter one. There is significant reduction in the width of the systematic uncertainties for the hotter star, even though all synthetic data have the same number of photons. The only difference between the two configurations is that, for the cooler star, the peak of the spectrum falls at the leftmost edge of the observed energy range while, for the hotter star, the observations sample well the rise-to and fall-from the spectral peak. This emphasizes the importance of encompassing the peak of the spectrum in the observed energy range to reduce the width of possible systematic errors.

\section{Summary}

Starting from an approximate light bending formula, we built an analytic model of neutron star thermal X-ray light curves that arise from two antipodal hot spots on its surface. We used this model to identify combinations of the physical model parameters that show weak correlations when fitting synthetic data, allowing for efficient sampling of the posterior distributions. We also followed this approach to demonstrate that the model parameters of interest (such as the neutron-star compactness) can be accurately constrained only when there is significant beaming in the emerging radiation from the surface. 

We considered neutron stars with effective temperatures $\lesssim 0.15\;$keV, as is the case for PSR~J0030$+$0451 and PSR~J0740+6620 observed with NICER, for which the peak of their X-ray spectrum falls on the edge (or outside) of the observed energy range. We showed that such small temperatures introduce additional systematics in the inference of the neutron-star compactness that can be as large as $\sim 0.05$. For a neutron star of mass $M=2.07~M_\odot$ and radius of $R=13$~km (as, e.g., had been inferred for PSR~J0740+6620), this systematic effect can be as large as $\delta R\sim (0.05/0.45)\times 13\sim 1.5$~km. 

\begin{acknowledgements}

    We thank Elizabeth Krause for useful discussions and suggestions on an early manuscript.

\end{acknowledgements}

\appendix

\section{Comparison of Pulse Profiles between numerical and analytical  models}\label{AppendixA}

In this Appendix, we compare the pulse profiles generated by the analytic model discussed in the main section of the paper to those calculated with a fully general-relativistic ray-tracing algorithm~\citep{Psaltis2014}. Our goal is to explore the range of parameters for which  the analytic model provides a good approximation for the pulse profiles from slowly spinning neutron stars. Even though we will focus on numerical profiles within the Schwarzschild+Doppler limit, i.e., we will not consider the effects of the spacetime spin and quadrupole moments, the numerical calculations include the effects of time delays and also depend on the size of the hotspot, which we assume to be infinitesimally small in the analytic model. Throughout this appendix, we will consider the configuration of a single circular hot spot on the stellar surface but ignore the effects of interstellar extinction and the response matrix of the detector. 

\begin{figure}[t]
	\centering
	\includegraphics[width=0.5\columnwidth, keepaspectratio]{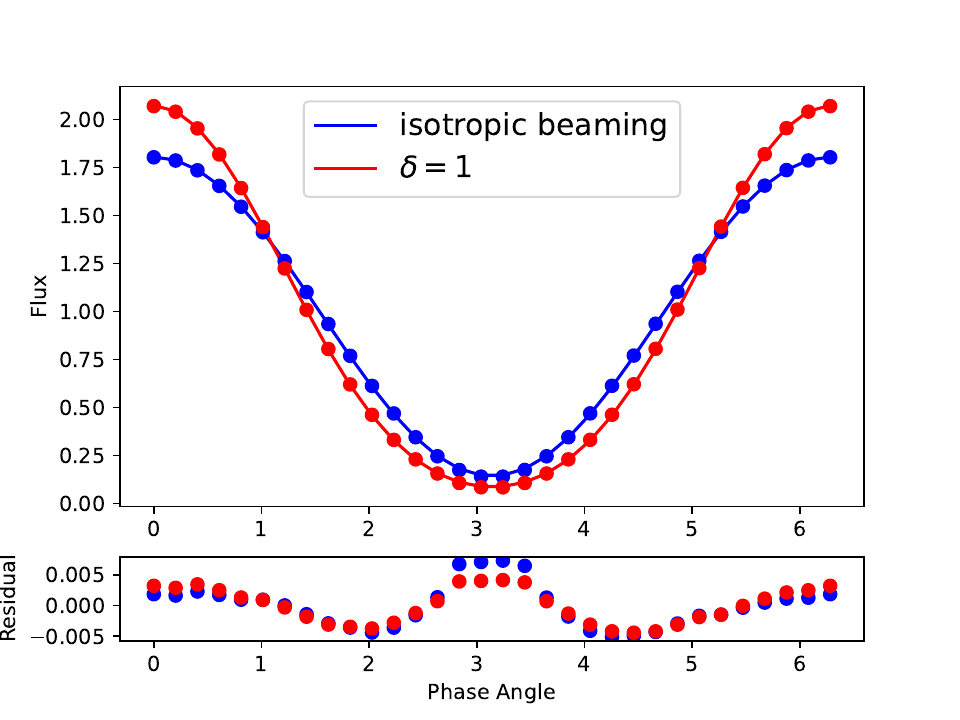}
	\caption{\label{fig:flux} Comparison of the pulse profiles from a small spot on a slowly rotating neutron star, generated using the analytic model (solid curves) and numerical ray-tracing calculations (filled circles) for two different beaming functions. The half opening angle of the spot is $5^\circ$, the spin frequency of the star is 1~Hz, its mass is $M=1.43~M_{\odot}$, its radius is $R=12$~km, and the inclination of the observer and spot colatitude are $\theta=65^\circ$ and $\zeta=50^\circ$, respectively. The flux is normalized by its mean value. The difference between the analytic model and the numerical result is less than 0.5\%, practically independent of the beaming assumed.}
\end{figure}

Figure~\ref{fig:flux} shows two sets of pulse profiles from a small spot (half opening angle of 5$^\circ$) on a slowly spinning ($f=1$~Hz) neutron star, with two different beaming functions. The solid curves are generated using the analytic model while the filled circles are the result of numerical ray tracing. The neutron star mass and radius are $M=1.43~M_{\odot}$ and $R=12$~km, respectively, corresponding to a compactness of $u=0.351$ in the analytic model. The inclination angle of the observer is set to $\theta=65^\circ$, the colatitude of the spot is $\zeta=50^\circ$, the observed photon energy is $0.325$~keV, and the effective temperature of the spot is $0.15$~keV. The flux is normalized by its mean value. For such a slowly spinning neutron star, the difference between the analytic model and the numerical result is less than 0.5\%, practically independent of the beaming assumed.

\begin{figure}[t]
	\centering
	\includegraphics[width=0.8\linewidth, keepaspectratio]{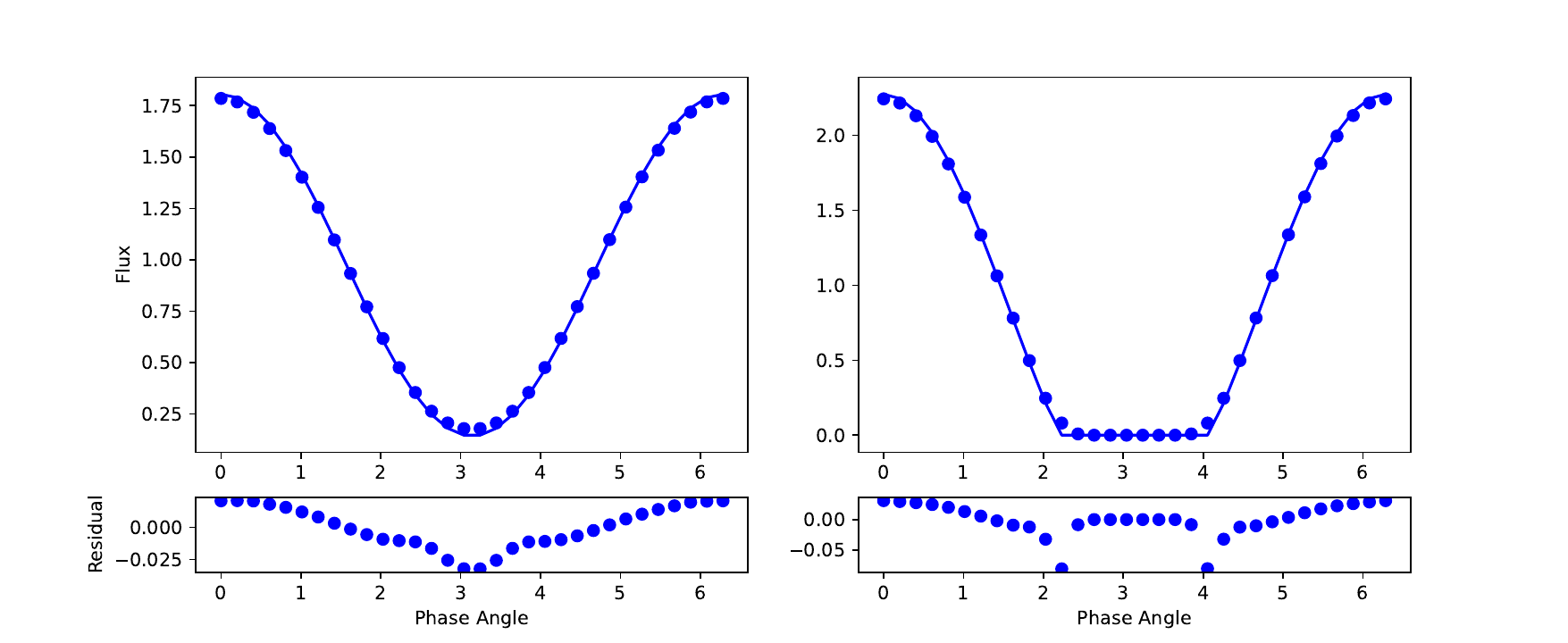}
	\caption{\label{fig:fluxr20} Comparison of the pulse profiles generated using the analytic model (solid curve) and numerical ray-tracing calculations (filled circles) when the half opening angle of the spot in the numerical model is set to $20^\circ$. In the left panel, the observer inclination and spot colatitude are set to $\theta=65^\circ$ and $\zeta=50^\circ$, making the hot spot visible throughout the pulse profile. In the right panel, these parameters are set to $\theta=80^\circ$ and $\zeta=80^\circ$, introducing an occultation of the spot for a fraction of the pulse profile. The remaining parameters are the same as in Figure~\ref{fig:flux}.}
\end{figure}

As a second set of tests, we explore the impact of the assumption in the analytic model of an infinitesimally small spot size. For these tests, we increased the half opening angle of the spot to $20^\circ$ in the numerical model, while keeping the other neutron-star parameters the same and an isotropic beaming function. We also explored two configurations in different visibility classes, one in which the spot is always visible and one in which it is occulted for a fraction of the pulse profile. Figure~\ref{fig:fluxr20} shows that increasing the spot size introduces only small ($\lesssim 2.5$\%) differences between the analytic and numerical results when the spot is visible throughout the pulse profile. However, for configurations in which the spot is occulted, the analytic model introduces a sudden truncation to the profile when the spot suddenly disappears, whereas in the numerical model the flux smoothly decreases to zero as the spot of finite size becomes increasingly occulted by the stellar surface. Even in this case, however, with an unrealistically large hot-spot size for the prime NICER targets, the difference between the two models for all but the times of ingress and egress remain at the percent level. Moreover, as discussed in the main text, for configurations with two antipodal (or nearly antipodal) hot spots, the occultation time of one spot coincides with the maximum-flux time of the other, which completely dominates the pulsed flux, reducing further the impact of this approximation (see \citealt{Baubock2015} for a detailed study of the effects of the hot-spot size on the pulse profiles).

\begin{figure}[t]
	\centering
	\includegraphics[width=0.8\linewidth, keepaspectratio]{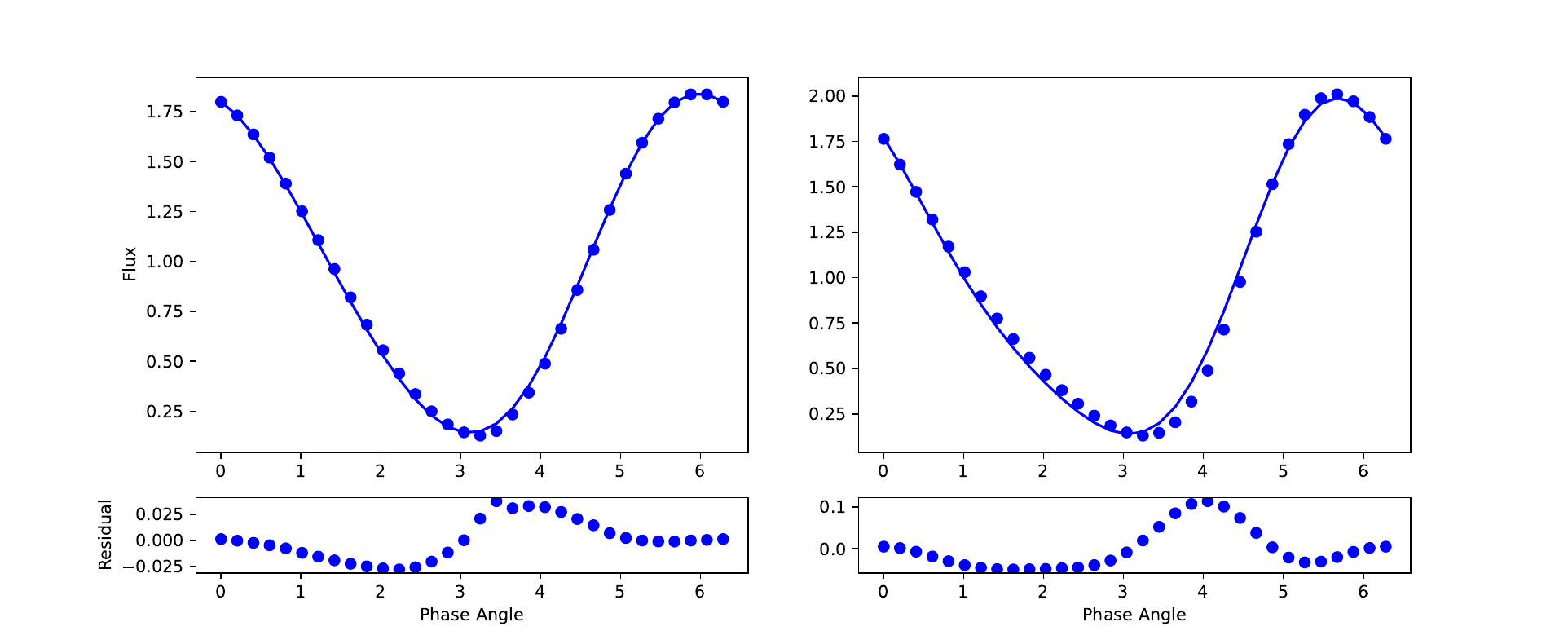}
	\caption{\label{fig:fluxf200} Same as in Fig.\ref{fig:flux} but for stellar spin frequencies of $f=200$~Hz in the left panel and $f=500$~Hz in the right panel. Ignoring the effects of time delays introduces a $\lesssim 2.5$\% error in the simulated pulse profiles for the slow spin frequencies of the primary NICER targets.}
\end{figure}

As a third set of tests, we explored the impact of ignoring time-delay effects in the analytic model. In this set, we changed the stellar spin frequency to 200~Hz and 500~Hz, kept the other parameters the same as in figure~\ref{fig:flux}, and assumed an isotropic beaming function. In this calculation with the analytic model, we took the Doppler factor into account instead of assuming $\delta_D=1$. Figure~\ref{fig:fluxf200} shows that, when the frequency of the neutron star is 200 Hz (left panel), the effect of neglecting time delays is $\lesssim 2.5$\% but, at 500 Hz (right panel), the effect can be as large as $\sim 10$\%. It is important to note, however, that for such fast spinning neutron stars, the higher moments of the spacetime become important and the neutron star is no longer spherical, introducing additional changes to the pulse profile that are of similar order~\citep{Psaltis2014}. Nevertheless, for the $\sim 170-270$~Hz spin frequencies of the primary NICER targets \citep{NICER1}, the effect of neglecting time delays is at the percent level.

From these results, we conclude that the use of the analytic model does not lead to any errors that impact the parameters inferences presented in this paper.

\section{An Approximate Form of Radiation Beaming From an Externally Bombarded Atmosphere}
\label{AppendixB}

In this Appendix, we motivate the functional form that we adopt for the beaming of radiation that emerges from the neutron-star atmosphere, following the spirit of the semi-analytic modeling approach followed throughout the paper.

We first employ the models of \citet[][see Fig.~8 of that paper]{atmosphere} and fit functions of the form
\begin{equation}
I(\mu)=I_0\left(1+h \mu\right)
\label{eq:beam_approx}
\end{equation}
to the calculated beaming functions, where $I(\mu)$ is the specific intensity at an angle with a cosine of $\mu$ with respect to the radial direction. As can be seen from that figure, equation~(\ref{eq:beam_approx}) is a good approximation for all but the smallest values of $\mu$, which have negligible contributions to the pulse profiles. 

We fit separately the resulting beaming functions at energies of 0.1, 0.3, 1, 3, and 10~keV. We also repeat the fit for three different power-law indices $\delta$ of the spectrum of the bombarding leptons. A shallow spectrum with $\delta=3$ contains a large fraction of low-energy leptons that deposit their energies at shallow depths and generate a flatter beaming function (smaller value of $h$). On the other hand, a steeper spectrum with $\delta=1$ contains a large fraction of high-energy leptons, which deposit their energies at large depths. In this case, the beaming function is centrally-peaked and very similar to the deep-heated atmospheres used in current NICER studies~\citep[see][for a discussion]{Baubock}.

The resulting values of the beaming index $h$ for each of these fits are shown in Figure~\ref{fig:beaming}. We then fit a quadratic to the beaming indices 
\begin{equation}
h(E,T_{\rm eff})=a+b\left(\frac{E}{kT_{\rm eff}}\right)+c\left(\frac{E}{kT_{\rm eff}}\right)^2\;,
\label{eq:beaming_energy}
\end{equation}
where we have normalized the photon energy to the effective temperature of the atmosphere ($T_{\rm eff}=0.4~$keV in this case). Table~\ref{table:fit} shows the best-fit parameters we obtained.

\begin{table}[t]
\caption{Beaming function parameters}
\begin{center}
\begin{tabular}{c c c c}
\toprule
$\delta$ & $a$ & $b$ & $c$ \\
\midrule
1 & 0.0500 & 0.3695 & -0.00976\\
2 & 0.0500 & 0.3106 & -0.00976\\
3 & 0.0500 & 0.0955 & -0.00976\\
\bottomrule
\end{tabular}
\end{center}
\label{table:fit}
\end{table}

Even though we have access to detailed radiation patterns for a set of simulations with a single effective temperature, we follow a set of arguments that motivate our use of equation~(\ref{eq:beaming_energy}) for other similar temperatures. 

The intensity of radiation emerging from an atmosphere at a photon energy $E=h\nu$ is given by the formal solution
\begin{equation}
I_\nu(\mu)=\int_{0}^{\infty} \left(\frac{1}{\mu}\right) S_\nu e^{-\tau_\nu/\mu}d\tau_\nu,
\end{equation}
where $S_\nu$ is the source function and $\tau_\nu$ is the optical depth. For pure absorption and emission, the source function is the equal to the blackbody intensity $I_{\rm b}(E,T)$ at the local temperature, i.e.,
\begin{equation}
I_\nu(\mu)=\int_{0}^{\infty} \left(\frac{1}{\mu}\right)I_b\left(E,kT\right) e^{-\tau_\nu/\mu}d\tau_\nu\;.
\end{equation}

Using our definition~(\ref{eq:beam_approx}) of the beaming factor, we write
\begin{equation}
h=\frac{1}{I_0}\frac{dI_\nu(\mu)}{d\mu}=\frac{1}{I_0} \left(-\frac{1}{\mu}I_\nu(\mu)
+\frac{1}{\mu^3}
\int_{0}^{\infty} I_b\left(E^\prime,kT^\prime\right) e^{-\tau_\nu/\mu}\tau_\nu d\tau_\nu
\right)\;.
\end{equation}
We now evaluate the integral in the above equations using integration-by-parts to obtain
\begin{equation}
    \frac{1}{\mu^3}\int_{0}^{\infty} I_b\left(E^\prime,kT^\prime\right) e^{-\tau_\nu/\mu}\tau d\tau_\nu=
    -\frac{1}{\mu}I_\nu(\mu)+
    \frac{1}{\mu^2}
    \int_{0}^{\infty} \frac{dI_b}{d\tau_\nu} e^{-\tau_\nu/\mu}\tau_\nu d\tau_\nu
\end{equation}
Combining these two equations and utilizing the fact that the intensity $I_0$ at the local normal to the surface is approximately equal to the blackbody intensity at the effective temperature of the atmosphere, i.e., $I_0\simeq I_b(E,T_{\rm eff})$, we find
\begin{equation}
h=\frac{1}{I_b(E,T_{eff}) \mu^2}\int_{0}^{\infty}e^{-\tau_\nu/\mu}\frac{dI_b}{d\tau_\nu}\tau_\nu d\tau_\nu\;.
\end{equation}
We have effectively converted the derivative with respect to the cosine of the beaming angle $\mu$ into a derivative of the blackbody function with respect to optical depth. Because the blackbody function depends on the local temperature in the atmosphere, we can then convert this into a derivative of the temperature with respect to optical depth.

We define $x\equiv h\nu/k T$ and $x_{eff}=h\nu/kT_{\rm eff}$ and write
\begin{equation}
h=\int_{0}^{\infty}\left(\frac{\tau_\nu}{\mu}\right) e^{-\tau_\nu/\mu}\frac{xe^x(e^{x_{\rm eff}}-1)}{(e^x-1)^2}\frac{1}{T}\frac{dT}{d\tau_\nu}d\left(\frac{\tau_\nu}{\mu}\right)\;.
\label{eq:h_1}
\end{equation}
To leading order, the local temperature for a deep heated atmosphere is related to the effective temperature in the atmosphere through the Eddington-Barbier relation~\citep{Dimitri}
\begin{equation}
T^4=\frac{3}{4}T_{eff}^4[\tau+q(\tau)],
\end{equation}
where $q(\tau)$ is the Hopf function and $\tau$ is an appropriate energy-averaged optical depth. This makes the $(1/T)(dT/d\tau_\nu)$ term in equation~(\ref{eq:h_1}) to be independent of $T$, to leading order. Furthermore, the factor $\tau_\nu/\mu e^{-\tau_\nu/\mu}$ is a function that sharply peaks at $\tau_\nu/\mu\simeq 1$ for all photon energies. As a result, to leading order, the integral scales with photon energy as
\begin{equation}
h\sim\frac{x_{\rm eff}e^{x_{\rm eff}}(e^{x_{\rm eff}}-1)}{4(e^{x_{\rm eff}}-1)^2}\;.
\end{equation}
According to this expression, the beaming factor $h$ is expected to depend primarily on the ratio $x_{\rm eff}=E/k T_{\rm eff}$ of the photon energy to the effective temperature of the atmosphere. In equation~(\ref{eq:beaming}) in the main text, we use this scaling to connect the beaming factors we have obtained here for the $T_{\rm eff}=0.4~$keV atmosphere to those of atmospheres with different effective temperatures. Figure~\ref{fig:beaming} compares the approximate scaling we derived here with the beaming factors that we obtained from fitting the numerical models, demonstrating the validity of our approximation.

\bibliographystyle{apj}
\bibliography{bib}

\end{document}